\documentclass[prd,showpacs,superscriptaddress,preprintnumbers,amsmath,nofootinbib]{revtex4}
 \usepackage[dvips,final]{graphicx}
  \usepackage{amssymb,amsmath,epsfig,bm,pifont}
\newcommand{\be}{\begin{equation}}\newcommand{\ee}{\end{equation}}%
\newcommand{\bd}{\begin{displaymath}}\newcommand{\ed}{\end{displaymath}}
\newcommand{\bit}{\begin{itemize}}                                        
 \newcommand{\eit}{\end{itemize}}                                         
\newcommand{\ben}{\begin{enumerate}}                                      
 \newcommand{\een}{\end{enumerate}}                                       
\newcommand{\baa}{\begin{array}{lll}}                                     
 \newcommand{\eaa}{\end{array}}                                           
\newcommand{\ba}{\begin{eqnarray}}                                        
 \newcommand{\ea}{\end{eqnarray}}                                         
\newcommand{\Ds}{\displaystyle}                                           
\newcommand{\gev}[1]{\relax\ifmmode{\text{GeV}^{#1}}                      
                     \else{GeV$^{#1}${ }}\fi}                             
\def\MSbar{\relax\ifmmode\overline                                        
            {\rm MS}\else{$\overline{\rm MS}${ }}\fi}                     
\def\as{\relax\ifmmode \alpha_s\else{$ \alpha_s${ }}\fi}                  
\def\abar{\relax\ifmmode{\bar{a}}\else{$\bar{a}${ }}\fi}                  
    
\usepackage{color}
\definecolor{mBlue}{rgb}{0,0,1}
 
\definecolor{mRed}{rgb}{1,0,0}
 
\definecolor{mGreen}{rgb}{0.0,1.0,1.0}
 
\begin{document}
\thispagestyle{empty}
 \date{\today}

\title{
Cut moments approach in the analysis of DIS data
}


\author{D. Kotlorz}
\email{dorota@theor.jinr.ru}
\affiliation{Institute of Mathematics and Physics,
                Opole University of Technology,\\
                45-758 Opole, Proszkowska 76, Poland}
\affiliation{Bogoliubov Laboratory of Theoretical Physics, JINR,\\
                141980 Dubna, Russia}
                                
\author{S. V. Mikhailov}
\email{mikhs@theor.jinr.ru}
\affiliation{Bogoliubov Laboratory of Theoretical Physics, JINR,\\
                141980 Dubna, Russia}

\author{O. V. Teryaev}
\email{teryaev@theor.jinr.ru}
\affiliation{Bogoliubov Laboratory of Theoretical Physics, JINR,\\
                141980 Dubna, Russia}
\author{A. Kotlorz}
\email{a.kotlorz@po.opole.pl}
\affiliation{Institute of Mathematics and Physics,
                Opole University of Technology,\\
                45-758 Opole, Proszkowska 76, Poland}

\begin{abstract}
We review the main results on the generalization of the DGLAP evolution
equations within the cut Mellin moments (CMM) approach,
which allows one to overcome the problem of kinematic constraints in 
Bjorken  $x$.
CMM obtained by multiple integrations as well as multiple differentiations
of the original parton distribution also satisfy the DGLAP equations with
the simply transformed evolution kernel. The CMM approach provides novel
tools to test QCD; here we present one of them.
Using appropriate classes of CMM, we construct the generalized Bjorken sum
rule that allows us to determine the Bjorken sum rule value  from the
experimental data in a restricted kinematic range of $x$.
We apply our analysis to COMPASS data on the spin structure function $g_1$.
\end{abstract}
\pdfinfo{%
  /Title    ()
  /Author   ()
  /Creator  ()
  /Producer ()
  /Subject  ()
  /Keywords ()
}

\maketitle
\section{Introduction}
\label{sec:intro}

In virtue of  QCD factorization in hard processes
 hadron properties in the deep inelastic scattering (DIS) can be described
in terms of the parton distribution functions (PDFs) $f_p(x,\mu^2)$.
They  are universal process-independent 
densities explaining how the whole hadron momentum $P$ is partitioned in $x\cdot P$
between partons of type $p$.
Here hard momentum transfer $q$:
$-q^2=Q^2 \gg P^2=m^2_h$, and the Bjorken variable $x$ satisfies $0< x = Q^2/(2Pq)<1$.
These distributions $f_p(x,\mu^2)$ are formed by nonperturbative strong interaction
at hadronic scale $m^2_h$,
while the dependence on the normalization scale $\mu^2$ is governed by the well-known
Dokshitzer-Gribov-Lipatov-Altarelli-Parisi (DGLAP) evolution equations
\cite{Gribov:1972ri,Gribov:1972rt,Dokshitzer:1977sg,Altarelli:1977zs} within perturbative QCD.
Alternatively, one can study how to evolve with this scale $\mu^2$ the Mellin
moments of the parton densities $f(n,\mu^2)$, which are integrals of PDFs weighted with
$x^n$ over the whole range (0,1) of $x$.
These moments provide a natural framework of QCD analysis as they
originate from the basic formalism of operator product expansion.
However,
these standard moments, in principle, cannot be extracted from any
experiment due to kinematic constraints inevitably
appearing in real DIS of lepton-hadron and hadron-hadron collisions.
 Namely, arbitrarily
small values of the variable $x$ cannot be reached in experiments,
which shows itself especially in ``fixed target'' experiments
like in JLab \cite{Deur:2008ej,Deur:2014vea}.
It would be useful to invent new ``real observables'' with a goal to overcome the kinematic constraints.
They were realized as the ``cut (truncated) Mellin
moments'' (CMM)
 $\Ds f(z;n,\mu^2)= \int^1_z f(x,\mu^2) x^{n-1} dx$, generalized moments of the parton
 distribution $f(x,\mu^2)$  in the
unavoidable lower limit of integration
 $z\equiv x_\text{min}=Q^2_\text{min}/(2(Pq)_\text{max}) > 0$, and in this
way the kinematic constraint can be taken into account.
This circumstance can be the main reason for large
uncertainties at data processing:
this effect is aggravated if a singularity of $f(x,\mu^2)$
in the neighborhood of $x=0$ is expected \cite{Deur:2014vea}.

The idea of ``truncated'' Mellin moments of the parton densities in QCD
analysis was introduced and developed in the late 1990s
\cite{Forte:1998nw,Forte:2000wh,Piccione:2001vf,Forte:2002us}.
The authors obtained the nondiagonal differential evolution
equations, in which the $n$th truncated moment couples to all higher ones.
Later on, diagonal integro-differential DGLAP-type evolution equations for
the single and double truncated moments of the parton densities were
derived in \cite{Kotlorz:2006dj} and \cite{Kotlorz:2009si,Kotlorz:2011pk}, respectively.
The main finding of the truncated CMM approach is that
the $n$th moment of the parton density also obeys the DGLAP equation,
but with a rescaled evolution kernel $P'(z)=z^n P(z)$ \cite{Kotlorz:2006dj}.
The CMM approach has already been successfully applied, e.g., in spin physics to
derive a generalization of the Wandzura-Wilczek relation in terms of the
truncated moments and to obtain the evolution equation for the structure
function $g_2$ \cite{Kotlorz:2011pk,Kotlorz:2014kfa}.
The advantages of the CMM approach to QCD factorization for DIS
structure functions were also presented in \cite{Kotlorz:2016icu}.
The truncation of the moments in the upper limit is less important in comparison
to the low-$x$ limit because of the rapid decrease of the parton densities as
$x\rightarrow 1$; nevertheless, a comprehensive theoretical analysis requires
an equal treatment of both truncated limits.
The evolution equations for double cut moments and their application to
study the quark-hadron duality were also discussed in \cite{Psaker:2008ju}.
Recently, a valuable generalization of the CMM approach incorporating
multiple integrations as well as multiple differentiations of the
original parton distribution has been obtained \cite{Kotlorz:2014fia}.
This novel generalization of CMM and the corresponding DGLAP equations
provides a powerful tool to test QCD at experimental constraints.
In Sec.~\ref{sec:sec2}, we briefly discuss the approach and present its main 
practically important results together with its DGLAP evolution.
Then we focus attention on a new important special CMM.
Based on this CMM, we construct in Sec.~\ref{sec:sec4}
a device to improve an experimental determination of the Bjorken polarized
sum rule.
In Sec.~\ref{sec:sec5}, we present the simplified form of the effective
method, based on the CMM, for practical use in analysis of data.
We apply it to the COMPASS measurements on $g_1$ \cite{Adolph:2015saz}
and also discuss the impact of the higher twist effects using Jlab data.

\section{CMM as solutions of DGLAP generalization}
\label{sec:sec2}
To apply our approach to specific cases of cut Mellin moments,
like the Bjorken polarized sum rule (BSR),
we consider it in more general context, as solutions of the DGLAP evolution \cite{Kotlorz:2014fia}.
Indeed, to deal with new distributions CMM to process DIS data, one should
know how the CMM can be evolved with the factorization scale $\mu^2$.
We review here a variety  of linear transformations $\hat{L}$ under the solutions of
the nonsinglet DGLAP
equation that lead to generalized CMM (gCMM) and then focus our attention on special cases of gCMM.
Suppose  $f(x,\mu^2)$ is a solution of the nonsinglet DGLAP equation with the kernel $P(y,a_s(\mu^2))$,
\begin{equation}\label{eq.2.1}
\dot{f}\equiv\frac{d }{d\ln \mu^2}f(z,\mu^2) =
(P\ast f)(z)\equiv
\int\limits_0^1 P(y,a_s(\mu^2))\;f(x,\mu^2)\;\delta(z-xy)\;dx\,dy,
\end{equation}
where the sign $\ast$ means Mellin convolution; the running coupling
$a_s=\alpha_s/(4\pi)$ satisfies the renormalization group equation with the QCD
$\beta$ function in the rhs
 $\Ds \mu^2 \frac{d}{d\mu^2}a_s(\mu^2)= -\beta\left(a_s(\mu^2)\right)$.
Then the linear transformed $f$, $f \to {\cal F}=\hat{L}f$,
 which is a generalization of CMM (see the second column of Table \ref{tb:1})
  is also the solution of the DGLAP equation:
\begin{subequations}
\label{sys.2.3}
\begin{equation}\label{eq.2.3}
\dot{{\cal F}} = ({\cal P}\ast {\cal F})
\end{equation}
with the kernel ${\cal P}$,
\begin{equation}\label{eq.2.4}
{\cal P}(y,a_s(\mu^2)) = \hat{L} P(y,a_s(\mu^2))\hat{L}^{-1},~\text{where}~\hat{L}\ast\hat{L}^{-1}=\delta(1-y)\,.
\end{equation}
 \end{subequations}
\begin{table}[h]
\begin{center}
\begin{tabular}{|c|c|c|} \hline
No.  & Generalized CMM ${\cal F}$                  & DGLAP Kernel ${\cal P}$\\  \hline
  &                                               &                  \\
1.&   $f(x)$                                      & $P(y)$           \\
 &                                               &                  \\
2.&   $x^nf(x)$                                   & $P(y)\cdot y^n$  \\
  &                                               &                  \\
3.&$f(z;n)=\int_{z}^{1}x^{n-1}\,f(x)\,dx$         & $P(y)\cdot y^n$  \\
  &                                               &                  \\
4.&   $f(z;\{n_i\}_k)=\int\limits_z^1 z_{k}^{n_k-1} dz_k
\int\limits_{z_{k}}^{1} z_{k-1}^{n_{k-1}-1} dz_{k-1}\: ...
\int\limits_{z_{2}}^{1} z_{1}^{n_1-1}\;f(z_1,\mu^2)\;dz_1\,$  & $\Ds P(y)\cdot  y^{ \sum_{i=1}^k n_i}$  \\
  &                                               &                  \\
5.&   $\Ds f(z;\{n,0\}_{\nu}) = \int_{z}^{1}\frac{\ln^{(\nu-1)}\left(x/z\right)}
{\Gamma(\nu)}~x^{n}f(x)\,\frac{dx}{x} $  & $P(y)\cdot y^n$  \\
  &                                               &                  \\
6.&   $\Ds f(z;\{n,1\}_{\nu}) = \int_{z}^{1}\frac{(x-z)^{\nu-1}}{\Gamma(\nu)}\,
~x^{n}f(x)\,\frac{dx}{x} $  & $P(y)\cdot y^{n+\nu-1}$            \\
  &                                               &                  \\
7.&   $\Ds -\frac{df(x)}{dx}$                         & $P(y)\cdot y^{-1}$     \\
  &                                               &                  \\
8.&$\Ds \left( -\frac{d}{dx}\right)^k\left[x^n f(x)\right]$& $P(y)\cdot y^{n-k}$\\
 &                                                & \\ \hline
  &                                                & \\
9.&   $\Ds f(z;* \omega) = (\omega \ast f)(z)$              & $P(y)$            \\
  &                                               &                  \\    \hline
\end{tabular}
\caption{\label{tb:1} Collection of the main results of CMM generalization of the DGLAP
equations.
The second  column contains the generalized CMM ${\cal F}$
 and the third column contains corresponding DGLAP
evolution kernels ${\cal P}$.}
\end{center}
\end{table}
\noindent
The different transformations $\hat{L}$ are presented in Table \ref{tb:1} explicitly:
in the second column---for ${\cal F}$, in the third one---for the corresponding DGLAP evolution kernel ${\cal P}$.
Item 4 lays the key role: all the other 
results below can be obtained from this ${\cal F}$.
They admit generalization from integer $k$ to
real $\nu$ for items 5, 6, and 8; see the discussion in \cite{Kotlorz:2014fia}.
The partial solutions in 7 and 8 were also considered earlier in \cite{Teryaev:2005uf,Artru:2008cp}.
The expression in item 5 admits differentiation and
integration with respect to the parameter $\nu$ and leads to new solutions.
The same is also true for the expression in item 6 with the evident additional modification
 of the kernel $P$ and the convolution in the right-hand side of the DGLAP equation.
Based on these gCMM, different interesting special solutions of the generalized DGLAP
equations (\ref{sys.2.3}) can be constructed and applied to an analysis of the experimental data.

It is evident that the singlet case keeps in force the same transformations
$\hat{L}$ under the quark $q(x,Q^2)$ and gluon $g(x,Q^2)$ distributions simultaneously and,
respectively, (\ref{eq.2.4}) under the matrix of the corresponding
 evolution kernels.
 In other words, Eq.~(\ref{eq.2.1}) can be extended to a homogeneous system of
 evolution equations together with symmetry transformations in Eq.~(\ref{sys.2.3}).

Now let us  focus on the transform in item 5 in Table~I.
The corresponding DGLAP kernel for it is independent of $\nu$.
Hence, integrands $\ln^{k}\left(x/z\right)/k!$ at different $k$
are ``bricks'' for any new gCMM constructions that evolve following the
DGLAP equation with the same kernel $P$.
Indeed, for \textit{any normalized weight} $\omega(t)$ the CMM $f(x;* \omega)$,
presented as a Mellin convolution of PDFs $f$ and $\omega$ (see item 9 of
Table \ref{tb:1}),
\begin{subequations}
 \label{eq.3.11}
 \begin{eqnarray}
 \label{eq.3.11a}
f(x) \to {\cal F}(x) \equiv f(x;* \omega)= \left(\omega \ast f\right)(x)&\equiv& \int_{x}^{1}\omega \left(x/z\right)
~f(z,\mu^2)\,\frac{dz}{z}, \\
&&\int_{0}^{1}\omega(t) dt =1\, ,
\end{eqnarray}
  \end{subequations}
 is normalized as $f$,
\begin{equation}\label{eq.3.12}
\int_{0}^{1}f(x;* \omega)\,dx = \int_{0}^{1}f(x)\, dx = 1.
\end{equation}
The corresponding DGLAP kernel ${\cal P}$ for the $f(x;* \omega)$ can be obtained directly
in virtue of the commutativity of
Mellin convolution,
${\cal P}=\omega \ast P \ast \omega^{-1} = P$~\footnote{
Notation $\omega^{-1}$ means that
$(\omega \ast \omega^{-1})(x)= (\omega^{-1} \ast \omega)(x)=\delta(1-x)$
or for the corresponding moments $\omega(n)$, $1/\omega(n)\cdot\omega(n) =1$.}.
The weight $\omega(t)$ can be considered as a result of appropriate (including infinite)
sums of the mentioned normalized bricks $\ln^{k}\left(t\right)/k!$;
each of them does not change the DGLAP kernel.
To return to the initial PDF $f(x)$, one must take $\omega(z)=\delta(1-z)$ in
the definition (\ref{eq.3.11}).
We shall investigate the applications of these properties for experimental data analysis
in the case of the nonsinglet spin structure function $g_1~$ (in other notation $g_1^{NS}$)
in the next sections.

\section{Generalized Bjorken sum rule}
 \label{sec:sec4}
We construct the generalized truncated moment $g_1(z,n,\omega)$
as a Mellin convolution of the function $g_1$ with any normalized
function $\omega(x)$, Eq.~(\ref{eq.3.11a}), which obeys the DGLAP
evolution equation with the rescaled kernel:
\begin{equation}\label{eq.4.2}
g_1(x,n;\omega) = \int_{x}^{1}
\omega \left(x/z\right)\,g_1(z)\,z^{n}\,\frac{dz}{z}\, ,
\end{equation}
\begin{equation}\label{eq.4.3}
{\cal P}(y) = P(y)\cdot y^{n}.
\end{equation}
For $n=0$ one obtains
\begin{equation}\label{eq.4.4}
g_1(x,0;\omega)= \left(\omega \ast g_1\right)(x)
\end{equation}
with the same evolution kernel as $g_1$, namely $P(y)$.
In this way, we define the cut Bjorken sum rules, $\Gamma_{1}(x_0)$,
 and simultaneously, the generalized cut Bjorken sum rules (gBSR),
$\Gamma_{1;\omega}(x_0)$,
\begin{eqnarray}
\Gamma_{1}(x_0) &=& \int_{x_0}^{1}g_1(x)\,dx\,,\label{eq.4.5b}\\
\Gamma_{1;\omega}(x_0) &=& \int_{x_0}^{1}g_1(x,0;\omega)\,dx, \label{eq.4.5a}
\end{eqnarray}
which are equal to the ordinary Bjorken sum rule as $x_0 \to 0$:
\begin{equation}\label{eq.4.6}
\Gamma_{1;\omega}(0)=\int_{0}^{1}g_1(x,0;\omega)\,dx
 = \int_{0}^{1} g_1(x)\, dx\equiv\Gamma_1(0).
\end{equation}
We shall  estimate the value of $\Gamma_1(0)$ from the smooth
extrapolation of the truncated moments $\Gamma_{1;\omega}(x_0)$ in $x_0$.
To this aim, we construct \textit{a bunch} of different $\Gamma_{1;\omega}(x_0)$.
Note that $\Gamma_{1;\omega}(x_0) \leqslant \Gamma_1(x_0)$ for any non-negative
$\omega$ that leads to one-side estimates $\Delta=\Gamma_1(x_0)-\Gamma_{1;\omega}(x_0)\geqslant 0$.
To extend the range of variation of the approach
and enable upper estimates of $\Gamma_1(x_0)$,
we construct a bunch of $\Gamma_{1;\omega}(x_0)$
 based on the simple sign-changing normalized function
$\omega (x)$ depending on three parameters $z_1,~z_2,~A$,
\begin{equation}\label{eq.4.9d}
\omega (z) = -A\,\delta(z-z_1)+(1+ A)\,\delta(z-z_2).
\end{equation}
Here the $\omega$-model parameters are $z_2 >z_1> x_0> 0$ and $A>0$ for the sign change.
This model, following (\ref{eq.4.4}), leads to a ``shuffle'' of the initial
PDF $g_1$ with different weights and arguments:
 \begin{eqnarray}
&& g_1(x,0;\omega)= -A\frac{\theta(z_1>x)}{z_1}g_1(x/z_1)+
(1+ A)\frac{\theta(z_2>x)}{z_2}g_1(x/z_2)\,, \\
&&\Gamma_{1;\omega}(x_0) = \int_{x_0/z_2}^{1}\,g_1(x)\,dx +
A\,\int_{x_0/z_2}^{x_0/z_1}\,g_1(x)\,dx. \label{eq.4.8}
\end{eqnarray}
The $\Gamma_{1;\omega}(x_0)$ approaches  $\Gamma_{1}(x_0)$ from above, $\Gamma_{1;\omega}(x_0) \geqslant \Gamma_1(x_0)$ for
\begin{equation}\label{eq.4.8a}
A > \int_{x_0}^{x_0/z_2}\,g_1(x)\,dx\,\Big/\int_{x_0/z_2}^{x_0/z_1}\,g_1(x)\,dx.
\end{equation}
We shall fit free $\omega$-model parameters in order to saturate the integral $\Gamma_{1;\omega}(x_0)$ as soon as possible when the parameter $x_0$ tends to $0$.
To this end, let us expand $\Gamma_{1;\omega}(0)$ into Taylor series around $x_0$,
\begin{equation}\label{eq.4.9}
\Gamma_1(0) = \Gamma_{1;\omega}(x_0 - x_0)=\Gamma_{1;\omega}(x_0) - x_0\,\Gamma'_{1;\omega}(x_0)
+ x_0^2\,\frac{1}{2}\,\Gamma''_{1;\omega}(x_0) +\cdots\, ,
\end{equation}
to estimate $\Gamma_1(0)$ in the lhs using a few first orders of Taylor expansion in the
rhs of Eq.~(\ref{eq.4.9}).
Requiring the first derivatives to vanish, $\Gamma'_{1;\omega}(x_0)=0$, or,
requiring the same  for the second one, $\Gamma''_{1;\omega}(x_0)=0$, to straighten the behavior of $\Gamma_{1;\omega}(x_0)$,
one can improve the approach to $\Gamma_1(0)$.

(1) Let us require $\Gamma'_{1;\omega}(x_0)=0$, then for the lhs of Eq.~(\ref{eq.4.9})
   one obtains the approximation:
\begin{equation}\label{eq.4.9c}
\Gamma_1(0) \approx \Gamma^{\rm{0APX}}_1(x_0) = \Gamma_{1;\omega}(x_0)+ 0+ \frac{1}{2}\,x_0^2\,\Gamma''_{1;\omega}(x_0).
\end{equation}
This condition fixes the value of the model parameter $A=A_{01}(x_0)$ and
then $\Gamma''_{1;\omega}(x_0)$:
\begin{eqnarray}
A_{01}(x_0)&=& \left[
\frac{t_1\,g_{1}(t_1)}{t_2\,g_{1}(t_2)}-1\right ]^{-1}\,, \label{eq.4.9b} \\
x_0^2\,\Gamma''_{1;\omega}(x_0) &=&
A_{01}(x_0)\,t_1^2\,g'_{1}(t_1) - \left[ 1+A_{01}(x_0)\right]t_2^2\,g'_{1}(t_2),
\label{eq.4.9e}
\end{eqnarray}
where here and below~$\Ds t_1=\frac{x_{0}}{z_1},~ t_2=\frac{x_{0}}{z_2}$.
For a special (single) root $x_0=x_{00}$ that satisfies
the condition
\begin{equation}\label{eq.4.9a}
\frac{1}{z_1}\, \frac{g'_{1}(t_1)}{g_{1}(t_1)} =
\frac{1}{z_2}\, \frac{g'_{1}(t_2)}{g_{1}(t_2)},
\end{equation}
the second derivation $\Gamma''_{1;\omega}(x_{00})$ vanishes also and the
approximation $\Gamma^{\rm{0APX}}_1(x_0)$ in (\ref{eq.4.9c}) in this case
reduces to
\begin{equation}\label{eq.4.12}
\Gamma_1(0) \approx \Gamma^{\rm{0APX}}_1(x_{00}) +0 +0
\end{equation}
with $A_{00}= A_{01}(x_{00})$. 

(2) Let us require now $\Gamma''_{1;\omega}(x_0)=0$, which leads to the first order
    approximation (IAPX),
 \label{eq.4.11}
\begin{equation}\label{eq.4.11a}
\Gamma_1(0) \approx \Gamma^{\rm{IAPX}}_1(x_0) =
\Gamma_{1;\omega}(x_0) - x_0\,\Gamma'_{1;\omega}(x_0)+0,
\end{equation}
with  $A=A_{02}(x_0)$ and $\Gamma'_{1;\omega}(x_0)$:
\begin{eqnarray}\label{eq.4.10}
A_{02}(x_0) &=& \left[
\frac{t_1^2\,g_{1}'(t_1)}{t_2^2\,g'_{1}(t_2)}-1\right ]^{-1}\,, \label{eq.4.13b}\\
x_0\,\Gamma'_{1;\omega}(x_0) &=&
A_{02}(x_0)\,t_1\,g_{1}(t_1) - \left[ 1+A_{02}(x_0)\right]t_2\,g_{1}(t_2).
\label{eq.4.13c}
\end{eqnarray}
To illustrate the features of $\Gamma_{1;\omega}$, we plot the bunch $\Gamma_{1;\omega}(x_0)$
in Eq.~(\ref{eq.4.8}) for different values of $A$ in Figs. 1 and 2, including:
``constant behavior'' value $A=A_{00}=A_{01}(x_{00})$ fixed at special root
$x_{00}\approx 0.037$,
``quasilinear behavior'' value $A=A_{02}(\bar{x})$ fixed at some value
$\bar{x}=0.01$ (\ref{eq.4.13b}), and the standard truncated Bjorken sum rule
$\Gamma_{1}(x_0)$, Eq.~(\ref{eq.4.5b}) (thick black curve).
\begin{figure}[ht]
\centering{
\begin{minipage}{0.49\textwidth}
\includegraphics[width=\textwidth]{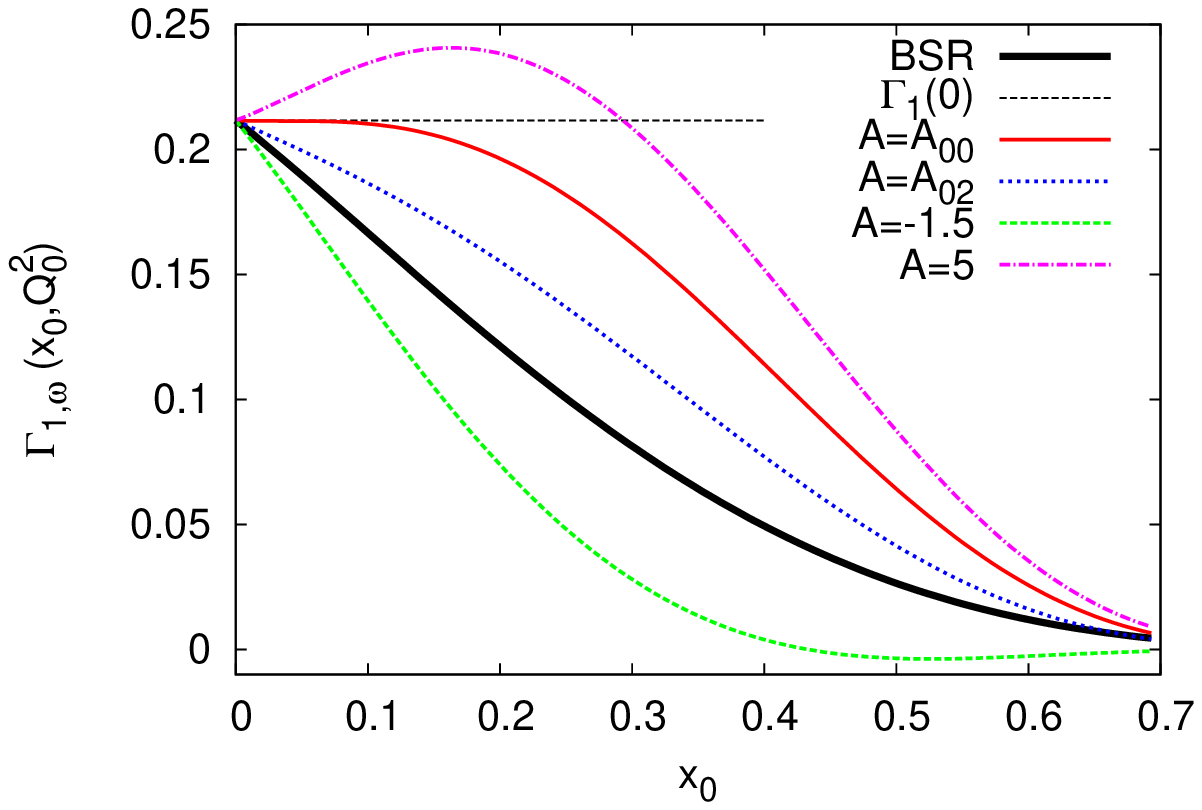}
\caption{\footnotesize \label{fig1}
$\Gamma_{1;\omega}(x_0)$, Eq.~(\ref{eq.4.8}), for different values of $A$
and the truncated BSR $\Gamma_{1}(x_0)$, Eq.~(\ref{eq.4.5b}) (thick black
curve) as a function of $x_0$. Input parametrization,
Eq.~(\ref{eq.4.13}), with $a=0$.}
\end{minipage}~~
\begin{minipage}{0.49\textwidth}
\includegraphics[width=\textwidth]{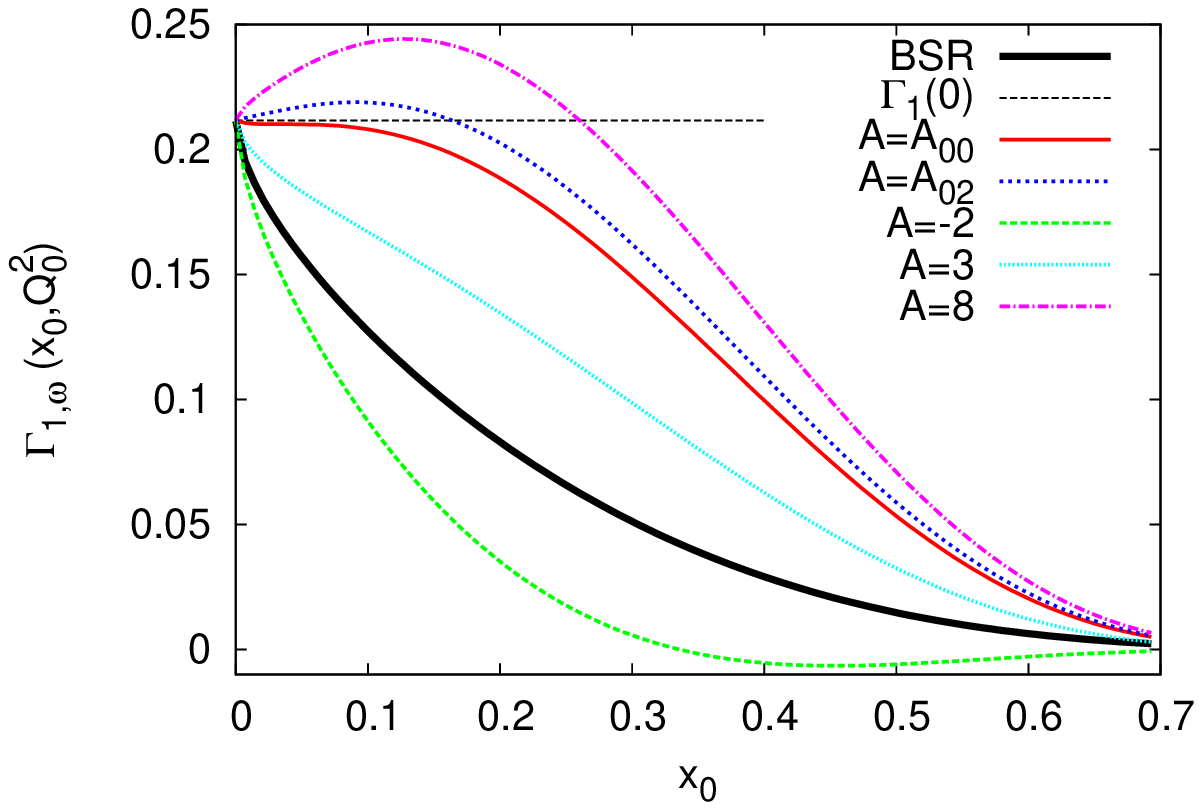}
\caption{\footnotesize \label{fig2}
$\Gamma_{1;\omega}(x_0)$, Eq.~(\ref{eq.4.8}), for different values of $A$
and the truncated BSR $\Gamma_{1}(x_0)$, Eq.~(\ref{eq.4.5b}) (thick black
curve) as a function of $x_0$. Input parametrization,
Eq.~(\ref{eq.4.13}), with $a=-0.4$.}
\end{minipage}}
\end{figure}
One can see that an appropriate model of $g_1$ shuffling can improve significantly
the approach $\Gamma_{1;\omega}(x_0)$ to $\Gamma_{1}(0)$; see, e.g., the red curve for
$A=A_{00}$.
The parameters of an optimal $\omega$ depend on the behavior of $g_1(x)$ (especially
in the neighborhood of zero),
which is fixed by different input parametrizations of $g_1$ at $Q_0^2=1\,\rm{GeV^2}$,
\begin{equation}\label{eq.4.13}
g_1(x,Q_0^2) = N\cdot x^a(1-x)^b(1+\gamma x),
\end{equation}
where $a=0$ in Fig.~1 and $a=-0.4$ in Fig.~2, respectively, at $b=3$,
$\gamma = 5$ and the coefficient $N$ is the norm.
In our tests, in order to obtain a smooth approach of the bunch in
the experimentally available $x$ region, we fixed $z_1=0.7$ and $z_2=0.9$.
The already mentioned  root $x_{00} \approx 0.037$ for the parametrization,
Eq.~(\ref{eq.4.13}) ($x_{00}$ value does not depend on the $a$ parameter of the
input), corresponds to approximation (\ref{eq.4.12}).
It is important to mention that the quasilinear regime near $0$ visibly starts at
rather large values of $x_0 \gtrsim 0.1$ for the different parametrization
in (\ref{eq.4.13}). This should ensue the applicability of approximation
(\ref{eq.4.11a})
even for JLab experimental conditions, where the admissible $x$ bunches are rather far from 0.
In practice, one can use fit to the data instead of the ready input
parametrization. It is worthy to notice that the analysis based on the
bunch behavior allows one to shift the available region of $x$ to smaller
values, $x_0=x\cdot z_2$. In this manner, using data from large $x$ and choosing
suitable values of $z_1$ and $z_2$, one is able to get an answer in a much smaller
$x$ region.\\

In this section, we have shown in detail how to construct the generalized
Bjorken sum rule and illustrated the mechanism of
shuffling in it. We have also presented different methods of estimation of
$\Gamma_1(0)$ within the gBSR approach. In the next section, we shall
present the simplified form of the most important equations of our approach,
rewritten in terms of experimental parameters, for practical use in
analysis of data.

\section{Practical analysis of data}
 \label{sec:sec5}
The generalized Bjorken sum rule enables one to analyze integrals over the experimentally
accessible $x$ range in a manner in which $\Gamma_{1;\omega}(x_0) > \Gamma_1(x_0)$.
In this way, for $x_0>0$, $\Gamma_{1;\omega}(x_0)$, Eq.~(\ref{eq.4.5a})
approaches $\Gamma_1(0)$ closer than the original BSR $\Gamma_1(x_0)$,
Eq.~(\ref{eq.4.5b}).
For practical purposes, we rewrite here the essential formulas from the previous
section in terms of experimental data and demonstrate the effective method
for the estimation of $\Gamma_1(0)$.
Thus, the gBSR, Eq.~(\ref{eq.4.8}), where the lower limit of integrations
has to be strictly related to the minimal $x$ accessible experimentally,
$x_{min}$, takes the form
\begin{equation}\label{eq.5.1}
\Gamma_{1;\omega}(x_{min},r) = \int_{x_{min}}^{1}\,g_1(x)\,dx +
A\,\int_{x_{min}}^{x_{min}/r}\,g_1(x)\,dx.
\end{equation}
The experimental lower value $x_{min}$ in the above equation is related to $x_0$
from Eq.~(\ref{eq.4.8}) via
$x_0=x_{min}\cdot z_2$. The ratio parameter, $r\equiv z_1/z_2$,
\begin{equation}\label{eq.5.2}
x_{min} \, < \, r \, < \, 1\, ,
\end{equation}
can also be chosen taking into account the set of experimental $x$ points.
Please note that in the above formulas $x_0$ and $z_2$ do not appear
separately, only as a ratio, $x_0/z_2=x_{min}$. It means that gBSR can mimic
a shift of the argument of the original BSR, $\Gamma_1(x_{min})$ to the smaller
one, $\Gamma_1(x_0)$.\\

\begin{figure}[h]
\begin{minipage}{0.49\textwidth}
\includegraphics[width=\textwidth]{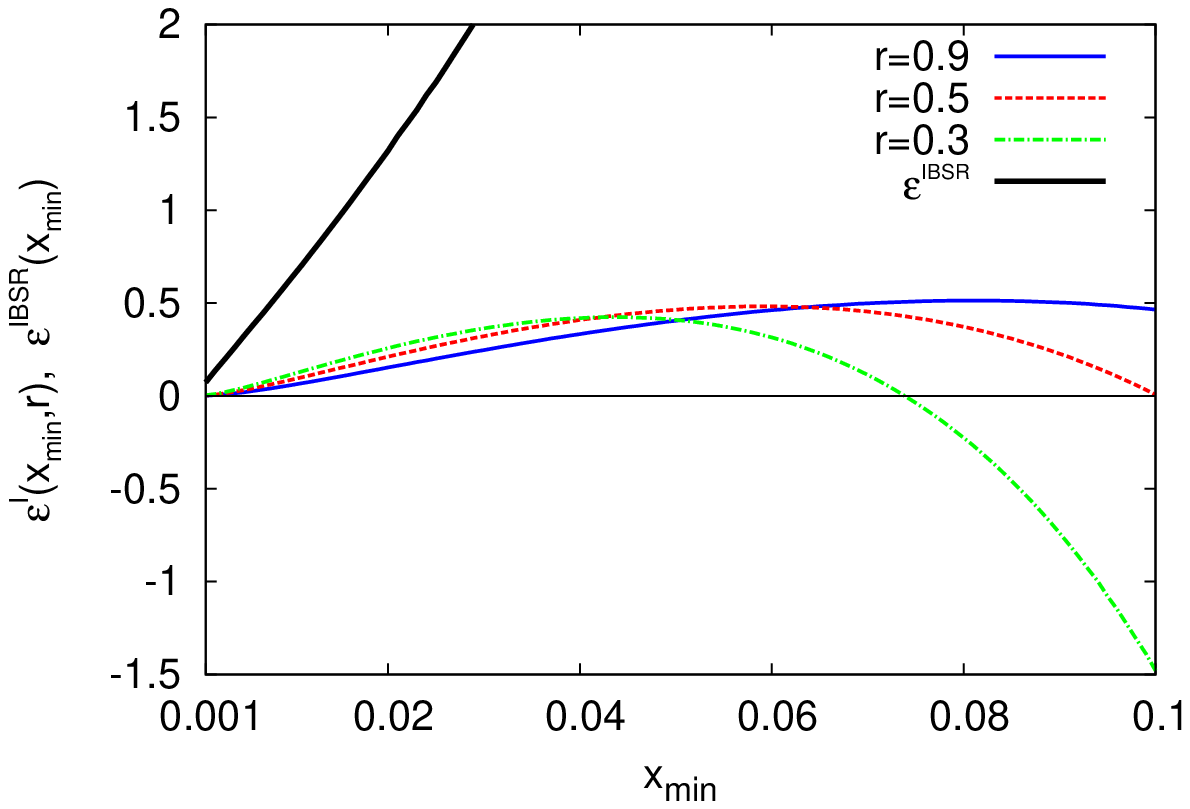}
\caption{\footnotesize
The percent errors $\epsilon^\text{I}(x_{min},r)$,
Eq.~(\ref{eq.5.6}), for different $r: 0.9, 0.5, 0.3$, together with
$\epsilon^\text{IBSR}(x_{min})$, Eq.~(\ref{eq.5.7}),
as a function of $x_{min}$.
Small-$x$ behavior of $g_1$ with $a=-0.1$, Eq.~(\ref{eq.4.13}).
\label{fig5new}}
\end{minipage}~~
\begin{minipage}{0.49\textwidth}
\includegraphics[width=\textwidth]{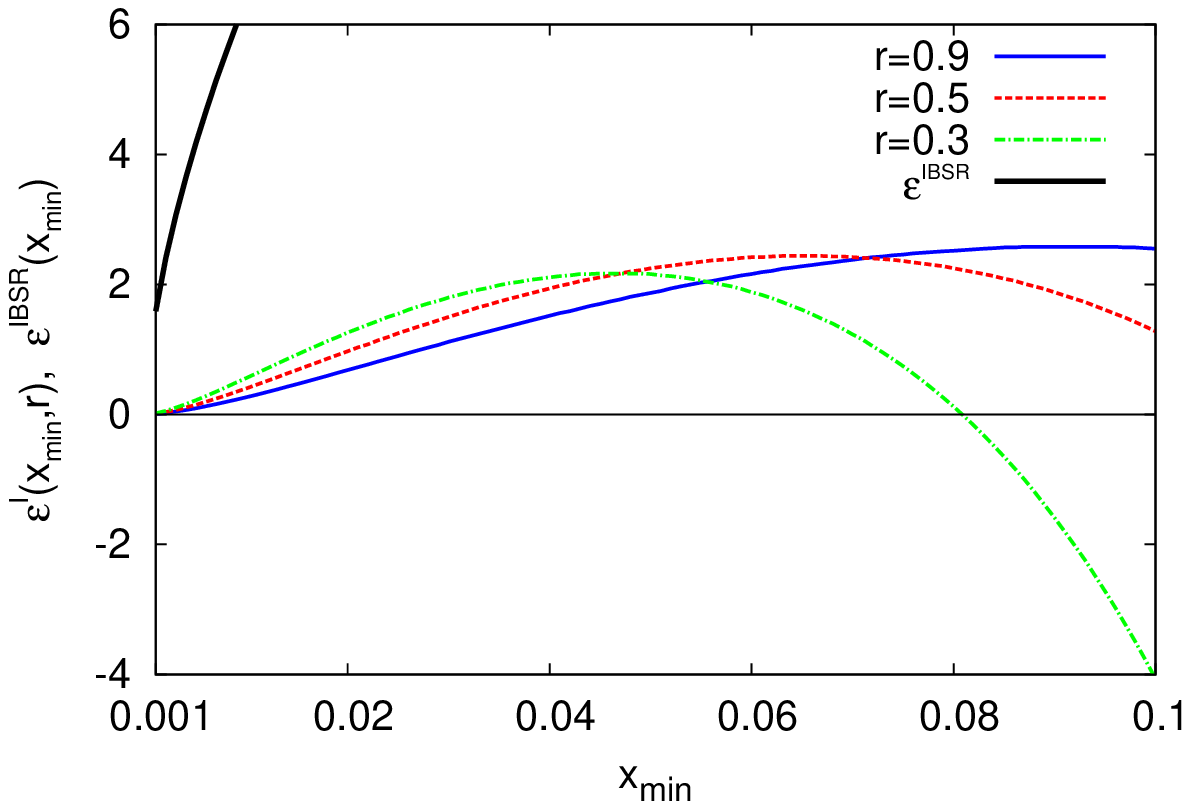}
\caption{\footnotesize
The percent errors $\epsilon^\text{I}(x_{min},r)$,
Eq.~(\ref{eq.5.6}), for different $r: 0.9, 0.5, 0.3$, together with
$\epsilon^\text{IBSR}(x_{min})$, Eq.~(\ref{eq.5.7}),
as a function of $x_{min}$.
Small-$x$ behavior of $g_1$ with $a=-0.4$, Eq.~(\ref{eq.4.13}).
\label{fig6new}}
\end{minipage}
\end{figure}

We have tested the methods of estimation of $\Gamma_1(0)$, described in
Sec.~\ref{sec:sec4} and have found that a very effective method, universal for
the different small-$x$ behavior of $g_1$ and for $x_{min}\lesssim 0.1$, is the
first order approximation, Eqs.~(\ref{eq.4.11a}) and (\ref{eq.4.10}).
With use of the experimental parameters $x_{min}$ and $r$, it reads
\begin{equation}\label{eq.5.3}
\Gamma_1(0) \approx \Gamma^{\rm{IAPX}}_1(x_{min},r) =
\Gamma_{1;\omega}(x_{min},r) + (A+1)\,x_{min}\, g_1(x_{min})
- A\frac{x_{min}}{r}g_1(x_{min}/r)
\end{equation}
with
\begin{equation}\label{eq.5.4}
A = \left[ r^2\,\frac{g'_1(x_{min}/r)}{g'_1(x_{min})}-1\right]^{-1}
\end{equation}
and $\Gamma_{1;\omega}(x_{min},r)$ given in Eq.~(\ref{eq.5.1}).
$\Gamma_1(0)$ from Eq.~(\ref{eq.5.3}) can be compared to the estimate from
the original BSR $\Gamma_1(x_{min})$, (\ref{eq.4.5b}), in the same first
order approximation,
\begin{equation}\label{eq.5.5}
\Gamma_1(0) \approx \Gamma^{\rm{IBSR}}_1(x_{min}) =
\Gamma_{1}(x_{min}) + x_{min}\, g_1(x_{min}).
\end{equation}
In Fig.~\ref{fig5new}, we plot the percent errors $\epsilon^\text{I}(x_{min},r)$,
\begin{equation}\label{eq.5.6}
\epsilon^\text{I}(x_{min},r) =
\left(\Gamma_1(0)-\Gamma^\text{IAPX}_1(x_{min},r)\right)/\Gamma_1(0)*100\%
\end{equation}
as a function of $x_{min}$ for three values of the ratio $r$. We assume a not
too singular small-$x$ behavior of $g_1$, $a=-0.1$ in Eq.~(\ref{eq.4.13}).
In Fig.~\ref{fig6new} we present the same but for a rather singular shape of $g_1$,
$a=-0.4$. For comparison, in both figures we show also the large error
$\epsilon^\text{IBSR}(x_{min})$,
\begin{equation}\label{eq.5.7}
\epsilon^\text{IBSR}(x_{min}) =
\left(\Gamma_1(0)-\Gamma^\text{IBSR}_1(x_{min})\right)/\Gamma_1(0)*100\%\, .
\end{equation}
The range of $x_{min}$ in our plots covers the smallest $x$ available in
the polarized experiments $\sim$ 0.004 at COMPASS, 0.02 at HERMES, and 0.1 at
Jlab. One can see a very good agreement of the estimated $\Gamma_1(0)$ with
its true leading order (LO) value (assuming $g_A/{g_V}=1.27$), for not too
singular behavior of $g_1$ at small $x$,
independently of the ratio $r$. For more singular behavior of $g_1$, this agreement
is still satisfactory and for $x_{min}\gtrsim 0.05$ it can be improved by taking
the ratio parameter $r$, Eq.~(\ref{eq.5.2}), as large as possible.

\begin{figure}[h]
\begin{minipage}{0.49\textwidth}
\includegraphics[width=\textwidth]{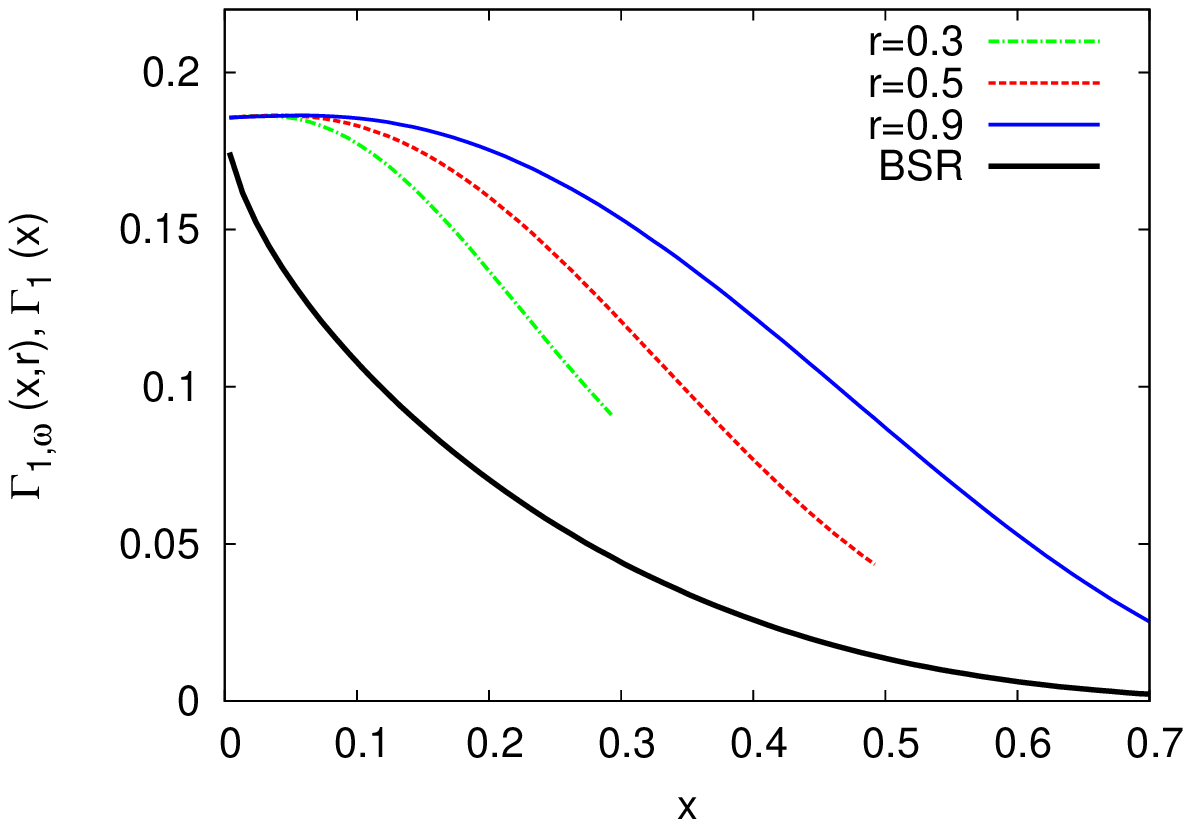}
\caption{\footnotesize
$\Gamma_{1,\omega}(x,r,Q^2)$, Eq.~(\ref{eq.5.1}), for
$A(x_{min}=0.0036,r)$, Eq.~(\ref{eq.5.4}) for three values of
$r: 0.9, 0.5, 0.3$, together with the truncated Bjorken sum rule
$\Gamma_{1}(x,Q^2)$, Eq.~(\ref{eq.4.5b}), as a function of $x$.
The results are based on our fit to the COMPASS data.
\label{fig7new}}
\end{minipage}~~
\begin{minipage}{0.49\textwidth}
\includegraphics[width=\textwidth]{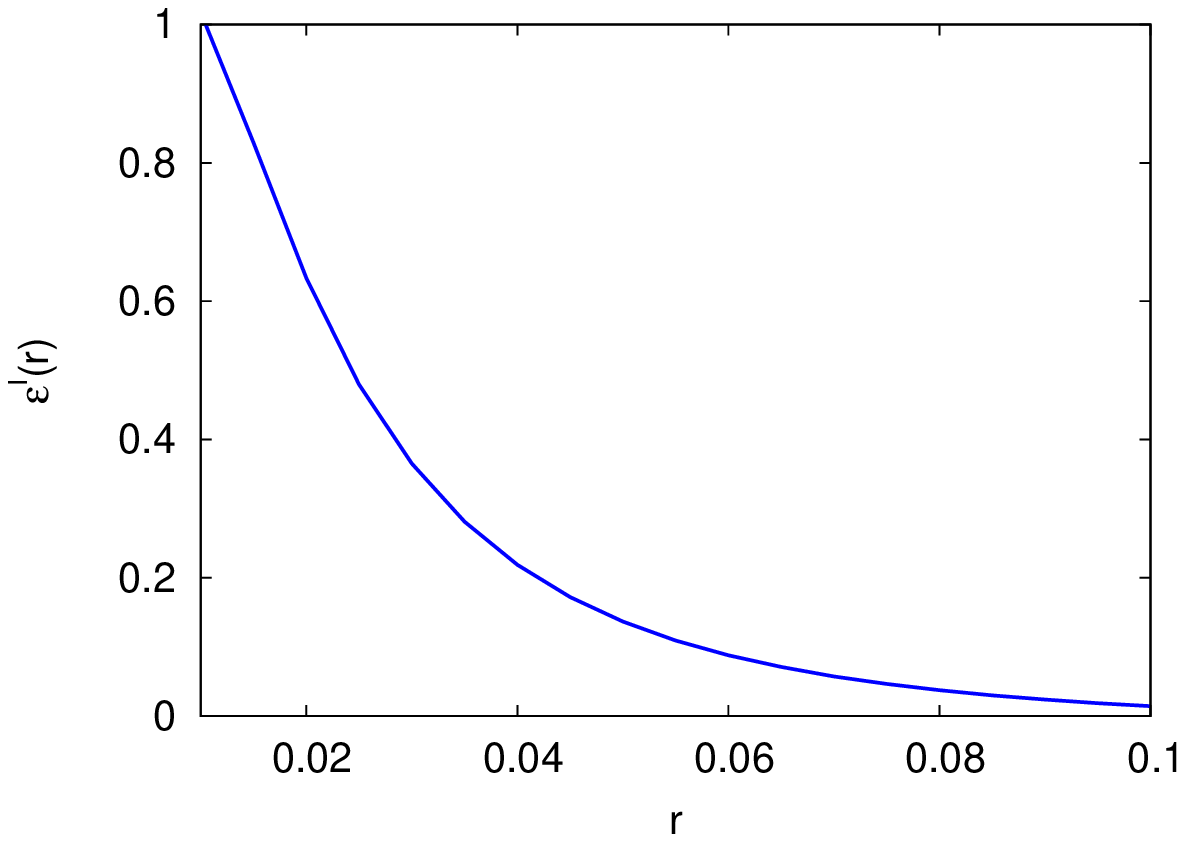}
\caption{\footnotesize
The percent errors $\epsilon^\text{I}(r)$,
Eq.~(\ref{eq.5.6}), for $x_{min}=0.0036$, as a~function of $r$.
The results are based on our fit to the COMPASS data.
\label{fig8new}}
\end{minipage}
\end{figure}

In Figs.~\ref{fig7new} and \ref{fig8new}, we present our results on determination
of the BSR based on the COMPASS \cite{Adolph:2015saz} data, where
$x_{min}=0.0036$.
We follow the method described above using
Eqs.~(\ref{eq.5.1})---(\ref{eq.5.4}).
We assume the input parametrization, Eq.~(\ref{eq.4.13}),
from our fit to the data at $Q^2=3$~\rm{GeV}$^2$:
\mbox{$g_1\sim x^{-0.42}(1-x)^{2.7}(1+3.4\,x)$.}\\
We find the following results for $x_{min}$ and different $r$:
$$
 \label{eq:COMPASS}
  \begin{array}{c|ccc|ccc}
\Gamma_1(0) & r & \Gamma_1^\text{IAPX} &~ \epsilon^\text{I}[\%]~ &
    r & \Gamma_1^\text{IAPX} &~ \epsilon^\text{I}[\%] \nonumber \\ \hline
0.186 &~ 0.1 & 0.185 & 1.4\cdot 10^{-2} &~ 0.9 & 0.186 & -7.7\cdot 10^{-3}
\end{array}
$$
One can see that for the first order $\Gamma^{\rm{IAPX}}_1(x_{min},r)$ approximation
the percentage error $\epsilon^\text{I}(x_{min}=0.0036)$, Eq.~(\ref{eq.5.6}),
is smaller than $1\%$ in the wide range $r>0.01$ and negligibly small for $r>0.05$.
These results, together with the accuracy estimates presented in
Figs.~\ref{fig5new} and \ref{fig6new}, confirm the efficiency of our integral
transform $\omega$ to estimate the BSR.

\begin{figure}[h]
\centering{
\begin{minipage}{0.48\textwidth}
\includegraphics[width=\textwidth]{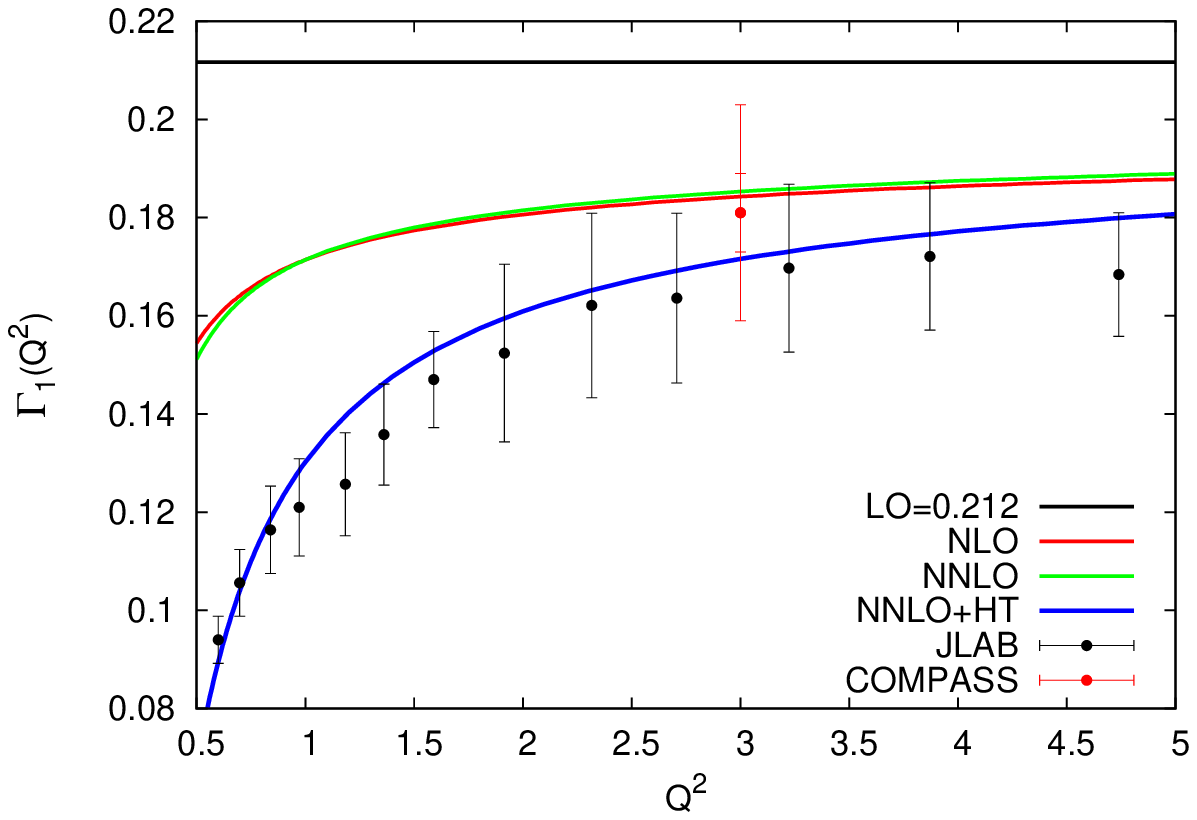}
\caption{\footnotesize \label{fig6}
$\Gamma_1(Q^2)$, Eq.~(\ref{eq.4.23}), incorporating \rm{N$^2$LO} and HT
corrections, $\mu_4^{p-n}/M^2=-0.047$, $\Lambda_\text{qcd} = 311$ MeV,
together with Jlab data \cite{Deur:2014vea}:
black point, the single red point higher is the COMPASS result
$0.181\pm 0.008\: {\rm stat.}\pm 0.014\: {\rm syst.}$ \cite{Adolph:2015saz},
$M$ is the nucleon mass.}
\end{minipage}
~~\begin{minipage}{0.48\textwidth} 
\includegraphics[width=\textwidth]{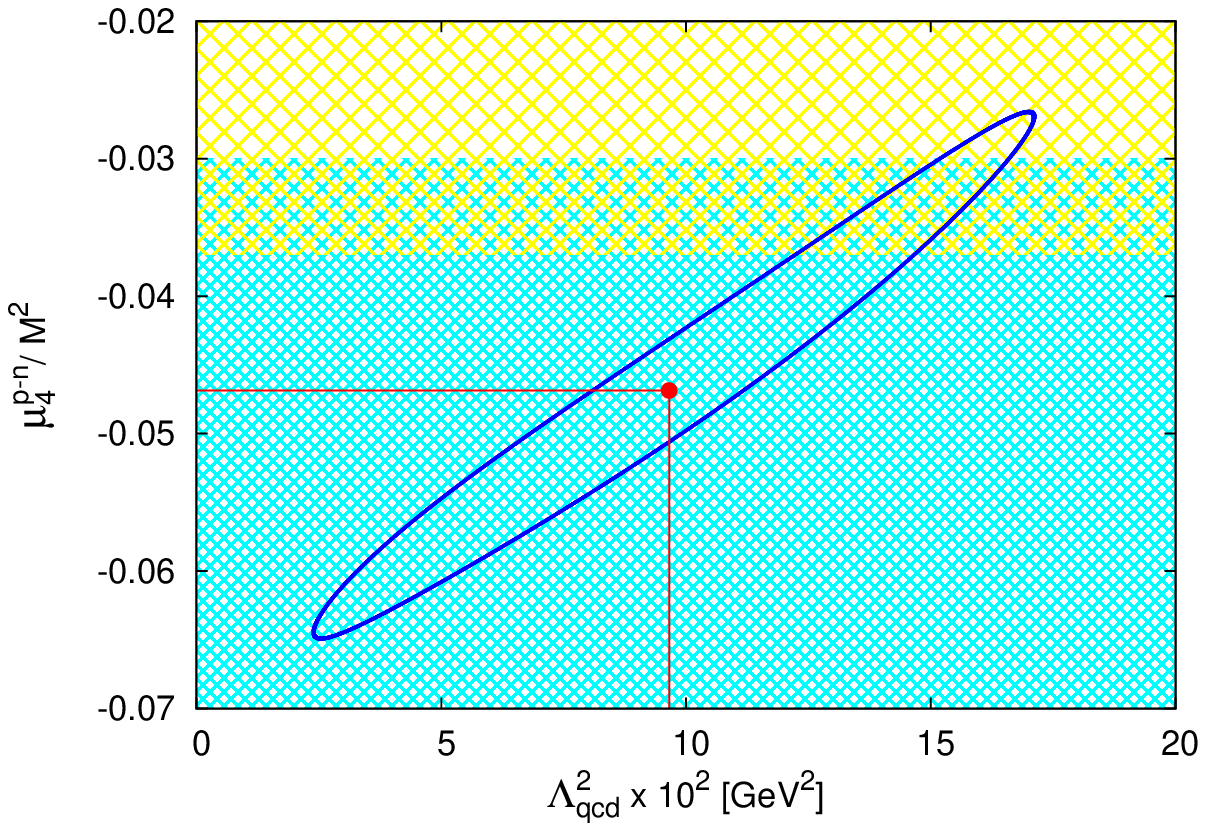}
\caption{\footnotesize \label{fig7}
Contour plot of 1$\sigma$ error ellipse for $\Lambda^2_\text{qcd}$ and
$\mu_4^{p-n}$, at the central point $\chi^2_\text{ndf} =0.60$.
The upper band (yellow strip) represents the Jlab result
$\mu_4^{p-n}/M^2 = -0.021\pm 0.016$ \cite{Deur:2014vea} and the lower band
(blue strip) is a typical theoretical estimation,
$\mu_4^{p-n}/M^2 = -0.05\pm 0.02$ \cite{Pasechnik:2009yc}.
}
\end{minipage}
}
\end{figure}

These estimates can be compared with the QCD result for the BSR
obtained in the \MSbar scheme in $O(\alpha_s^n)$, $n=1, 2, 3$ and $4$ approximation
in \cite{Kodaira:1978sh,Gorishnii:1985xm,Larin:1991tj} and
\cite{Baikov:2010je}, respectively, and incorporating higher twist (HT) effects,
\begin{equation}\label{eq.4.23}
\Gamma_1(Q^2) =
\frac{1}{6}\,\frac{g_A}{g_V}\left[\, 1-\frac{\alpha_s}{\pi}
-3.58\left(\frac{\alpha_s}{\pi}\right)^2
-20.22\left(\frac{\alpha_s}{\pi}\right)^3 -
175.7\left(\frac{\alpha_s}{\pi}\right)^4 \right] + \frac{\mu_4^{p-n}}{Q^2}\,.
\end{equation}
Here $\alpha_s \equiv \alpha_s(Q^2)$ is the running QCD coupling,
the coefficients of expansion are taken for the number of active quarks $n_f=3$,
and $\mu_4^{p-n}$ is the scale of the first power correction to the HT.
The HT effects become essential in the small/moderate $Q^2$ region;
see the analysis of its  impact for BSR in \cite{Pasechnik:2009yc}.
In our analysis $Q^2$ is of the order of a few $\rm{GeV}^2$ and
the HT impact is visible, which is shown in Fig.~\ref{fig6}.

To illustrate the reasonableness of the new estimates for $\Gamma_1(Q^2)$,
we have processed the JLab results \cite{Deur:2014vea} following
Eq.~(\ref{eq.4.23}) taken at N$^2$LO,
i.e., holding the first three terms in the perturbation part there.
The results of the fit are shown in Figs.~\ref{fig6} and \ref{fig7}.
In Fig.~\ref{fig7}, we present 1$\sigma$ error ellipse for two adjusted fit parameters:
$\Lambda_\text{qcd} = 311\pm{}^{103}_{156}$~MeV
and HT $\mu_4^{p-n}/M^2 =  -0.047\pm{}^{0.020}_{0.018}$; $M$ is the nucleon mass.
These values look reasonable in view of the actual world average data:
$\Lambda_\text{qcd} = 332\pm 17$~MeV \cite{Olive:2016xmw}
and $\mu_4^{p-n}/M^2 = -0.05\pm 0.02$ \cite{Pasechnik:2009yc}, \cite{Deur:2014vea}.

\section{Conclusions}
\label{sec:concl}
The QCD analysis of real data for the deep inelastic scattering processes faces
the principal problem: Bjorken variable $x$ is constrained by the unavoidable
kinematic condition (from below)
$x \geqslant x_\text{min}=Q^2_\text{min}/(2(Pq)_\text{max}) > 0$.
This is important for data processing, especially for the case of PDF
$f_p(x,\mu^2)$ increasing as $x \to 0$.
The CMM approach has been elaborated  just to overcome
this problem.
In this paper, we have reviewed the main results of the CMM
approach and suggested its generalization
that allows one to study the fundamental integral characteristics of the
parton distributions in an experimentally restricted region of $x$.
We demonstrated how, with the help of the so-called generalized Bjorken sum
rule, one can determine the BSR $\Gamma_1$ from experimental data in
the available  $x$ region.
We applied our approach to the COMPASS data and obtained good agreement with
the QCD predictions for the BSR, incorporating higher twist effects
estimated from the Jlab measurements.
Concluding, the presented method seems to be promising in the analysis of the
QCD sum rules.

\acknowledgments
This work is supported by the Bogoliubov-Infeld Program, Grant No.
01-3-1113-2014/2018.
S.V.M. acknowledges support from the BelRFFR-JINR, Grant No. F16D-004.

\bibliography{refs}
\bibliographystyle{apsrev}

\end{document}